\documentclass[12pt,dvipdfmx]{article}
\usepackage[letterpaper, left=.8in, top=0.9in, right=.8in, bottom=0.70in,nohead,includefoot, verbose, ignoremp]{geometry}
\usepackage{charter} 
\usepackage{enumerate} 
\usepackage{latexsym,amssymb,amsmath,amsfonts,graphicx,color,amsthm,enumerate} 
\usepackage[svgnames,dvipsnames,x11names]{xcolor}

\usepackage[round]{natbib}
\usepackage{bibentry}
\nobibliography*

\usepackage{bm}
\usepackage{booktabs}
\usepackage{multirow}

\theoremstyle{plain}
\newtheorem*{thm*}{Theorem}
\theoremstyle{definition}
\newtheorem*{dfn*}{Definition}
\newtheorem*{lem*}{Lemma}

\begin{document}
\vspace{-1in}
\title{On the Proofs of the Predictive Synthesis Formula}
\author{
Riku Masuda\footnote{INTAGE, Inc.
\newline{
E-Mail: rikumasuda520@gmail.com}} \
and 
Kaoru Irie\footnote{Corresponding author. Faculty of Economics, The University of Tokyo. 
\newline{
E-Mail: irie@e.u-tokyo.ac.jp}}
}
\maketitle
\begin{abstract}
Bayesian predictive synthesis is useful in synthesizing multiple predictive distributions coherently. However, the proof for the fundamental equation of the synthesized predictive density has been missing. In this technical report, we review the series of research on predictive synthesis, then fill the gap between the known results and the equation used in modern applications. We provide two proofs and clarify the structure of predictive synthesis. 
\par\vspace{4mm}
\end{abstract}

\section{Introduction}

Suppose that a decision maker (DM) is interested in predicting quantity $y$. This DM has $K$ agents (statistical models, human experts, deterministic computer algorithms, etc) who report their forecasts of $y$ or related quantities. Specifically, for each $k\in \{ 1,\dots ,K\}$, the $k$-th agent provides the DM with forecast density function $h_k$. The DM has her/his own prior for $y$, denoted by $p(y)$. In this context, the DM is interested in updating prior $p(y)$ to the posterior, conditioning information set $\mathcal{H} \equiv \{ h_1,\dots ,h_K \}$. However, the formal Bayes rule is not available unless the DM fully specifies her/his belief on $y$ and $\mathcal{H}$ jointly, which is often a difficult task in practice.

Thus, a coherent rule of updating the DM's forecast of $y$ based on the agents' opinion $\mathcal{H}$, other than the direct posterior computation by the Bayes rule, is needed. It is claimed that such a rule must be of the form
\begin{equation} \label{eq:bps}
    p(y|\mathcal{H}) = \int \alpha (y|\bm{x}) \prod _{k=1}^K h_k(x_k) d\bm{x}, 
\end{equation}
where $\bm{x} = (x_1,\dots ,x_K)$ and $\alpha (y|\bm{x})$ is a probability density function. This model, known as {\it Bayesian predictive synthesis}, has been utilized in modern forecasting problems, including the predictions of macroeconomic variables such as GDP and inflation rates \citep{McAlinnEtAl2017,mcalinn2021mixed,chernis2023predictive,chernis2024combining,chernis2024decision}, oil prices \citep{aastveit2023quantifying}, returns from financial assets \citep{tallman2024bayesian,tallman2024predictive}, prices of real estates \citep{cabel2022spatially}, and public health data \citep{sugasawa2023bayesian,kobayashi2024predicting,babasaki2024ensemble}. The comparison with other methods for model synthesis, such as stacking, logarithmic pooling and discounting, has also been investigated \citep{McAlinnEtAldiscussionBA2018,carvalho2022bayesian,bernaciak2024loss}. Examples of synthesis density $\alpha(y|\bm{x})$ are provided and studied in \cite{JohnsonWest2022}. Once the synthesis density is specified, the DM's forecast can be computed by the Markov chain Monte Carlo or sequential Monte Carlo methods \citep{masuda2023sequential}. 

Equation~(\ref{eq:bps}) appeared in \cite{mcalinn2019dynamic} (or \citealt{mcalinn2017dynamic}) for the first time, referring \cite{westcrosse1992} as its origin. However, the proof of (\ref{eq:bps}) cannot be found in \cite{westcrosse1992} and the reference therein, including \cite{genest1985modeling} and \cite{West1988,West1992d}. In this technical report, we try to fill the gap between the foundational results on predictive synthesis obtained in 1980s-90s and equation~(\ref{eq:bps}) used today as the Bayesian predictive synthesis formula. We review the known results in Section~2, then prove equation~(\ref{eq:bps}) in two ways in Section~3.

\section{Review of established results}

The original work of \cite{genest1985modeling} starts their analysis from the coherent synthesis of a single agent ($K=1$) who forecasts the occurrence of a single event. Their result has been extended by \cite{westcrosse1992} to the case of receiving the forecasts of multiple events from the agent. Equation~(\ref{eq:bps}) with $K=1$ is obtained as the limit when the number of events increases. The predictive synthesis of multiple agents has been discussed in \cite{genest1985modeling}, but limited to the single event case. The question is whether the same extension to the multiple events is feasible, so that (\ref{eq:bps}) is proven for general $K\ge 1$. Table~\ref{tab:papers} summarizes those various scenarios considered in the literature.

Before proceeding to the proof of (\ref{eq:bps}), we review the results established in the two papers.

\begin{table}[h]
\centering
\begin{tabular}{ ccc } 
 & Single event & Multiple events \\ \toprule 
 Single agent   & \cite{genest1985modeling} & \multirow{2}{*}{\cite{westcrosse1992}} \\ 
 $(K=1)$ &  \cite{westcrosse1992} & \\ \midrule 
 Multiple agents & \multirow{2}{*}{\cite{genest1985modeling}}   & \multirow{2}{*}{This study} \\ 
 $(K>1)$ & & \\ \bottomrule  
 \hline
\end{tabular}
\caption{Scenarios of predictive synthesis} \label{tab:papers}
\end{table}

\subsection{Single agent, single event}

Suppose that the DM's interest is in the occurrence of an event, $E \subset \Omega$. The DM's prior for $E$ is $p\equiv \mathbb{P}(E) \in [0,1]$. There is an agent who assesses the probability that event $E_1$, which is not necessarily identical to $E$, will happen at $f_1 \in [0,1]$. The DM is supposed to have a prior for $f_1$, or a probability distribution on $[0,1]$. For simplicity, we assume the existence of a density function $p(f_1)df_1$. Since we (and the DM) are unable, or reluctant, to specify the joint model of $E$ and $f_1$, the conditional probability, $\mathbb{P}(E|f_1)$, is not computable by the Bayes rule. Then, our goal is to obtain another formula that defines the probability of $E$ given $f_1$, denoted by $p^{\ast}(f_1)$. We require this probability to satisfy the {\it Consistency Condition}, which makes the joint, marginal and conditional probabilities compatible.

\begin{dfn*} [Consistency Condition]
For any $p$ and $p(f_1)$, the updating rule $p^{\ast}(f_1)$ must satisfy
\begin{equation*}
    p = \int _0^1 p^{\ast}(f_1) p(f_1) df_1,
\end{equation*}
and $p^{\ast}(f_1) = \mathbb{P}(E|f_1)$, in addition to other initial specifications.
\end{dfn*}
``Other initial specifications'' are typically the condition for the (marginal) prior moments of $f_1$. Here, we focus on the case of partially specified prior with mean $\mu_1$. 
\begin{dfn*} [Partially specified prior]
Prior $p(f_1)$ is partially specified if the DM does not fully specifies $p(f_1)$ but its mean by $\mu _1$. That is, 
\begin{equation*}
    \mu _1 = \int _0^1 f_1 p(f_1) df_1.
\end{equation*}
\end{dfn*}
Thus, for given $\mu_1$, the Consistency Condition must be satisfied for any prior with mean $\mu_1$. 
\cite{genest1985modeling} showed in their Theorem~2.1 that, under the Consistency Condition, the $p^{\ast}(f_1)$ must be of the form,
\begin{equation*}
    p^{\ast} (f_1) = p + \lambda _1(f_1 - \mu_1),
\end{equation*}
where $\lambda _1$ is a constant and contained in some finite interval so that $p^{\ast}(f_1)\in [0,1]$. \cite{westcrosse1992} pointed out in their Corollary~1 that this expression can be rewritten as,
\begin{equation} \label{eq:avg}
    p^{\ast} (f_1) = \mathbb{P}(E|f_1=1) f_1 + \mathbb{P}(E|f_1=0) (1-f_1). 
\end{equation}
To evaluate $p^{\ast}(f_1)$ in practice, the DM must specify either of the following sets of quantities:
\begin{itemize}
    \item $p\equiv \mathbb{P}(E)$ and $\mu_1$: DM's prior event probability and expected agent's forecast.
    
    \item $\mathbb{P}(E|f_1=1)$ and $\mathbb{P}(E|f_1=0)$: event probabilities when extreme forecasts are reported.
\end{itemize}

\subsection{Single agent, multiple events} \label{sec:single_multi}

\cite{westcrosse1992} extended the above result to the case where the agent assesses the probabilities of $n$ events, $E_1,\dots , E_n$, that are mutually disjoint, while the DM is interested in another event, $E$, with prior $p\equiv \mathbb{P}(E)$. Let $E_{n+1} \equiv ( \cup E_i )^c$. Denote the agent's forecast probabilities of $E_i$ by $f_i$ for each $i\in \{ 1,\dots, n, n{+}1 \}$. Note that $f_i\ge 0$ for all $i$ and $\sum _{i=1}^{n+1} f_i =1$, hence $\bm{f}\equiv (f_1,\dots ,f_n, f_{n+1})'$ is on the $n$-simplex. The DM has a prior for $\bm{f}$, denoted by density $p(\bm{f})$, whose marginal density is $p(f_i)$. The DM specifies its marginal mean by $\mu_i$: 
\begin{equation*}
    \mu _i = \int _0^1 f_i \ p(f_i) \ df_i, \ \ \ \ \ \ i\in \{ 1,\dots, n, n+1 \} 
\end{equation*}
Then, the Consistency Condition determines the DM's probability of $E$ given $\bm{f}$ by,  
\begin{equation} \label{eq:linear_pool}
    p^{\ast}(\bm{f}) = p + \sum _{i=1}^n \lambda_i (f_i - \mu _i), 
\end{equation}
or, equivalently, 
\begin{equation} \label{eq:wc1}
    p^{\ast}(\bm{f}) = \sum _{i=1}^{n+1} \Pi_i f_i, \ \ \ \ \ \mathrm{where} \ \ \ \ \Pi_i \equiv \mathbb{P}(E|f_i=1).
\end{equation}
Note that $(n{+}1)$-th event, or the complement of the $n$ events, is included in the latter expression of (\ref{eq:wc1}). The $(n{+}1)$-th term, $\Pi_{n+1}f_{n+1}$, is the ``residual,'' or the variation of $y$ that is unexplained by the forecasts of $n$ events, serving as the intercept term in (\ref{eq:bps}) as a part of $\alpha (y|x)$. 

The synthesis formula in (\ref{eq:bps}) is obtained for $K=1$ as follows. Let quantity of interest be $Y$. This is a random variable whose probability law represents the DM's uncertainty about this quantity. For arbitrary value $y$, set $E=\{ Y\le y \}$. In this context, the prior probability of event $E$ is the DM's distribution function of $Y$, or $p(y) \equiv \mathbb{P}(Y\le y)$. The agent forecasts quantity $X$, which is not necessarily equal to $Y$. For arbitrary points $-\infty = q_0<q_1<\cdots <q_n<q_{n+1}=\infty$, set $E_i = \{ q_{i-1}<X\le q_i\}$. Denote the agent's distribution and density function of $X$ by $H$ and $h$, respectively. Then the forecast made by the agent is written as $f_i = H(q_{i+1}) - H(q_i)$ for each $i$. This is the agent's probability that $X$ falls in interval $(q_{i-1},q_i]$. The DM has a prior for $\bm{f}$, but specifies only the expectation of $f_i$ as $\mu _i = \mathbb{E}[ f_i ] = \mathbb{E}[ H(q_i) - H(q_{i-1}) ]$ for each $i$. 

Let $\mathcal{H}_n \equiv \{ H(q_i) | i=1,\dots ,n  \}$. When being conditioned, this information set is equivalent to $\{ f_1,\dots, f_n \}$. Then equation (\ref{eq:wc1}) becomes, 
\begin{equation*} 
    \mathbb{P}( Y\le y | \mathcal{H}_n ) = \sum _{i=1}^{n+1} \Pi_i(y) \{ H(q_i) - H(q_{i-1}) \}, \ \ \ \ \ \mathrm{where} \ \ \ \ \Pi_i(y) = \mathbb{P}( Y\le y | H(q_i) - H(q_{i-1}) = 1 ).
\end{equation*}
Now, consider the equal-length grids by setting $q_i-q_{i-1} = \epsilon _n > 0$ for $i\in \{ 2,\dots ,n\}$, and suppose that $\epsilon _n\to 0$ as $n\to \infty$. Then, the expression above can be viewed as the Riemann sum and has the limit: 
\begin{equation*} 
    \mathbb{P}( Y\le y | \mathcal{H}_n ) = \sum _{i=1}^{n+1} \Pi_i(y) \frac{ H(q_i) - H(q_{i-1}) }{ \epsilon _n } \epsilon_n \to  \int \Pi (y|x) h(x) dx = \mathbb{P}( Y\le y | \mathcal{H} ),
\end{equation*}
where $x$ is the point at which intervals degenerate in the limit. 
The symbols in the right are defined as the limit of those in the left, as explained below. 
\begin{itemize}
    \item $\mathcal{H} = \lim_{n\to\infty} \mathcal{H}_n$. This information set contains almost all values of $H(x)$. Hence, having information set $\mathcal{H}$ is equivalent to knowing density function $h$. 
    
    \item $\Pi (y|x) =  \lim_{n\to\infty} \Pi _i(y) = \lim_{n\to\infty} \mathbb{P}(Y\le y|H(q_i)-H(q_{i-1})=1)$. This is the limit of the DM's distribution function when the agent reports that $X$ falls in interval $(q_{i-1},q_i]$. Note that this interval degenerates at point $x$ as $\epsilon _n \to 0$. Thus,  $\Pi (y|x)$ is the DM's distribution function when the agent is 100\% sure that $X=x$. 
\end{itemize}
Computing the derivatives of both sides in $y$ yields
\begin{equation*} 
    p( y | \mathcal{H} ) = \int \alpha (y|x) h(x) dx, \qquad \mathrm{where} \quad  \alpha (y|x) = \frac{d}{dy} \Pi(y|x).
\end{equation*}
This is equivalent to (\ref{eq:bps}) when $K=1$.

\subsection{Multiple agents, multiple events}

We now assume $K\ge 1$. Each of $K$ agents forecasts $n$ events. For $k\in \{ 1,\dots , K\}$, these events are denoted by $E_{k,1} , \cdots , E_{k,n}$ and are mutually disjoint. Then, the $k$-th agent assesses the probability that $E_{k,i_k}$ will happen at $f_{k,i_k}\in [0,1]$ for $i_k \in \{ 1,\dots,n \}$, while satisfying $\sum _{i_k=1}^n f_{k,i_k} \le 1$. Set $E_{k,n+1} \equiv (\cup _{i_k} E_{k,i_k})^c$ and $f_{k,n+1} \equiv 1 - \sum_{i_k = 1}^n f_{i_k,k}$. Let $\bm{f}_k \equiv (f_{k,1},\dots ,f_{k,n},f_{k,n+1})'$ and $\bm{f}_{1:K} \equiv \{ \bm{f}_1,\dots,\bm{f}_K \}$. Note that $\bm{f}_k$ is $(n{+}1)$-dimensional and includes $f_{k,n+1}$, hence sums to unity. 

The DM has a joint prior for $\bm{f}_{1:K}$, but does not fully specify its distributional form. Only their marginal means are specified as $\mu_{k,i_k} = \mathbb{E}[f_{k,i_k}]$ for each $k$ and $i_k$. The DM is interested in the occurrence of event $E$ and has a prior probability $p\equiv \mathbb{P}(E)$. After hearing the agents' opinions $\bm{f}_{1:K}$, the DM updates her/his belief on $E$ to $p^{\ast}(\bm{f}_{1:K})$ so that the Consistency Condition is satisfied.

Similarly to (\ref{eq:linear_pool}), for any partially specified prior $p(\bm{f}_{1:K})$, the $p^{\ast}(\bm{f}_{1:K})$ must be of the form,
\begin{equation} \label{eq:linear_pool_multi}
    p^{\ast}(\bm{f}_{1:K}) = p 
    + \sum _{{i_1}=1}^n \lambda_{1,i_1} (f_{1,i_1} - \mu _{1,i_1})
    + \cdots 
    + \sum _{{i_K}=1}^n \lambda_{K,i_K} (f_{K,i_K} - \mu _{K,i_K}) 
    .
\end{equation}
The coefficient, $\lambda _{k,i_k}$, must satisfy many inequalities so that $p^{\ast}(\bm{f}_{1:K}) \in [0,1]$. \cite{genest1985modeling} proved this expression for $n=1$ in their Theorem~3.2.

The proof of (\ref{eq:linear_pool_multi}) completely parallels to that of Lemma~2 in \cite{westcrosse1992}. That is, we first apply Lemma~3.1 in \cite{genest1985modeling} to obtain the form of (\ref{eq:linear_pool_multi}), treating this problem as the ``single agent, $nK$ events'' case without the sum-to-unity condition for each $\bm{f}_k$. Then restrict the support of the DM's prior for $\bm{f}_{1:K}$ to the product of $K$ simplices. Such a prior exists; a typical example is the independent Dirichlet distributions.

\section{Proof of general synthesis equation}

Our goal is to prove by using (\ref{eq:linear_pool_multi}) that, for any partially specified prior $p(\bm{f}_{1:K})$, we have 
\begin{equation} \label{eq:discrete_bps}
    p^{\ast}(\bm{f}_{1:K}) = \sum_{i_1 = 1}^{n+1} \cdots \sum _{i_k=1}^{n+1} \Pi_{(i_1,\dots, i_K)}
    \prod_{k = 1}^K f_{k,i_k}, 
\end{equation}
where the coefficient is the conditional probability:
\begin{equation*}
    \Pi_{(i_1,\dots , i_K)} \equiv \mathbb{P}\left( E \left| f_{1,i_1} = 1 , \cdots, f_{K,i_K} = 1 \right. \right).
\end{equation*}
Clearly, this is the multiple agents version of (\ref{eq:avg}). Similarly to the single agent case, this form is viewed as the Riemann sum that converges to (\ref{eq:bps}). Thus, to obtain equation~(\ref{eq:bps}), we need to prove (\ref{eq:discrete_bps}). We provides two approaches to the proof, each of which clarifies the mechanism of predictive synthesis in different ways.

\subsection{Proof by marginalization}

Our first proof explicitly addresses the update of the DM's information. Mathematically, it is the proof by induction. Note that this equation has been shown for $K=1$ in (\ref{eq:wc1}). Suppose that the DM initially has $K{-}1$ agents ($K\ge 2$), then adds another agent to her/his information set. The DM's prior must be consistent in the sense that we should be able to recover $p(\bm{f}_{1:K-1})$ by marginalizing $\bm{f}_K$ in the prior $p(\bm{f}_{1:K})$. 

Using (\ref{eq:linear_pool_multi}) for each $k$ individually, we have the DM's probability conditional on $k$ agents as, 
\begin{equation}\label{eq:linear_pool_k} 
    p^{\ast}(\bm{f}_{1:k}) = p + \bm{\lambda}_{1|k}' (\bm{f}_1 - \bm{\mu}_1) + \cdots + \bm{\lambda}_{k|k}' (\bm{f}_{k} - \bm{\mu}_{k}), 
\end{equation}
where we set $\bm{\lambda}_{j|k} \equiv (\lambda_{j,1|k} , \cdots , \lambda_{j,n+1|k})'$ and $\bm{\mu}_j \equiv (\mu _{j,1},\dots ,\mu _{j,n+1})'$ for $j\in \{ 1,\dots ,k\}$. By construction, coefficient $\lambda _{j,i_j|k}$ might be dependent on $k$, or the number of agents in the DM's information set. Note that $\lambda _{j,n+1|k} = 0$, $f_{j,n+1} = 1-f_{j,1}-\cdots -f_{j,n}$ and $\mu _{j,n+1} = 1 - \mu_{j,1}-\cdots - \mu_{j,n}$. Under this notation, if $f_{j,n+1}=1$, then $f_{j,1}=\cdots = f_{j,n}=0$ and 
\begin{equation*}
    \bm{\lambda}_{j|k}' (\bm{f}_j - \bm{\mu}_j) = -\lambda_{j,1|k} \mu_{j,1} - \cdots - \lambda_{j,n|k} \mu_{j,n} = -\bm{\lambda}_{j|k}' \bm{\mu}_j. 
\end{equation*}

Next, we define $\Pi _k$ for each $k$ as, 
\begin{equation*}
    \Pi_k \equiv \mathbb{P}(E|f_{1,n+1} = 1, \cdots , f_{k,n+1} = 1) = p - \bm{\lambda}_{1|k}' \bm{\mu}_{1} - \cdots - \bm{\lambda}_{k|k}' \bm{\mu}_{k},
\end{equation*}
where the second equality is obtained by setting $f_{1,n+1} = 1, \cdots , f_{k,n+1} = 1$ in the expression of $p^{\ast}(\bm{f}_{1:k})$ in (\ref{eq:linear_pool_k}). 
Then, observe the following relationship: 
\begin{equation*}
    \Pi_k = \Pi_{k-1} + (\bm{\lambda}_{1|k-1} - \bm{\lambda}_{1|k})'\bm{\mu}_{1} + \cdots 
     + (\bm{\lambda}_{k-1|k-1} - \bm{\lambda}_{k-1|k})' \bm{\mu}_{k-1}
    - \bm{\lambda}_{k|k} ' \bm{\mu}_{k}.
\end{equation*}
Similarly, by setting $f_{1,i_1} = 1 , \cdots , f_{k,i_k} = 1$ in (\ref{eq:linear_pool_k}), we have
\begin{equation*}
    \mathbb{P}(E|f_{1,i_1} = 1 , \cdots , f_{k,i_k} = 1) = \Pi_k + \lambda_{1,i_1|k} + \cdots + \lambda_{k,i_k|k}.
\end{equation*}
Using these notations and relationships, we obtain the following identities:
\begin{equation*}
    \begin{split}
    \sum_{i_k = 1}^{n+1} \mathbb{P}(E|f_{1,i_1} = 1 , \cdots , f_{k,i_k} = 1)f_{k,i_k} &= \sum_{i_k = 1}^{n+1} [\Pi_k + \lambda_{1,i_1|k} + \cdots + \lambda_{k,i_k|k} ]\  f_{k,i_k} \\
    &= [\Pi_k + \lambda_{1,i_1|k} + \cdots + \lambda_{k-1,i_{k-1}|k} ] + \bm{\lambda}_{k|k}'\bm{f}_k.
    \end{split}
\end{equation*}
Writing $\Pi_{k}$ by using $\Pi_{k-1}$, then using $\mathbb{P}(E|f_{1,i_1} = 1 , \cdots , f_{k-1,i_{k-1}} = 1)$, we have 
\begin{equation}\label{eq:key}
    \begin{split}
    \sum_{i_k = 1}^{n+1} \mathbb{P}(E|&f_{1,i_1} = 1 , \cdots , f_{k,i_k} = 1)f_{k,i_k} = \mathbb{P}(E|f_{1,i_1} = 1 , \cdots , f_{k-1,i_{k-1}} = 1) \\
    & \ \ \ \ + (\bm{\lambda}_{1|k-1} - \bm{\lambda}_{1|k})'\bm{\mu}_{1} + \cdots 
    + (\bm{\lambda}_{k-1|k-1} - \bm{\lambda}_{k-1|k})'\bm{\mu}_{k-1} + \bm{\lambda}_{k|k} ' (\bm{f}_{k} - \bm{\mu}_{k}) \\
    & \ \ \ \ + (\lambda_{1,i_1|k} - \lambda_{1,i_1|k-1}) + \cdots + ( \lambda_{k-1,i_k|k}-\lambda_{k-1,i_k|k-1} ).
    \end{split}
\end{equation}
Now, assume that (\ref{eq:discrete_bps}) holds for $K{-}1$: 
\begin{equation*}
    p^{\ast}(\bm{f}_{1:K-1}) = \sum_{i_1 = 1}^{n+1} \cdots \sum_{i_{K-1} = 1}^{n+1}  \mathbb{P}(E|f_{1,i_1} = 1 , \cdots , f_{K-1,i_{K-1}} = 1) f_{1,i_1} \cdots f_{K-1,i_{K-1}}.
\end{equation*}
Then, we confirm that (\ref{eq:discrete_bps}) also holds for $K$ as follows: 
\begin{equation*}
    \begin{split}
    & \sum_{i_1 = 1}^{n+1} \cdots \sum_{i_K = 1}^{n+1}  \mathbb{P}(E|f_{1,i_1} = 1 , \cdots , f_{K,i_K} = 1) f_{1,i_1} \cdots f_{K,i_K} \\
    &= \sum_{i_1 = 1}^{n+1} \cdots \sum_{i_{K-1} = 1}^{n+1}f_{1,i_1} \cdots f_{K-1,i_{K-1}} \sum_{i_K = 1}^{n+1} \mathbb{P}(E|f_{1,i_1} = 1 , \cdots , f_{K,i_K} = 1)f_{K,i_K} \\
    &= \sum_{i_1 = 1}^{n+1} \cdots \sum_{i_{K-1} = 1}^{n+1}  \mathbb{P}(E|f_{1,i_1} = 1 , \cdots , f_{K-1,i_{K-1}} = 1) f_{1,i_1} \cdots f_{K-1,i_{K-1}} \\
    &\ \ \ \ + (\bm{\lambda}_{1|K-1} - \bm{\lambda}_{1|K})'\bm{\mu}_{1} + \cdots +(\bm{\lambda}_{K-1|K-1} - \bm{\lambda}_{K-1|K})' \bm{\mu}_{K-1}
    + \bm{\lambda}_{K|K} ' (\bm{f}_{K} - \bm{\mu}_{K}) \\ 
    &\ \ \ \ + \sum_{i_1 = 1}^{n+1} f_{1,i_1} (\lambda_{1,i_1|K} - \lambda_{1,i_1|K-1}) + \cdots + \sum_{i_1 = 1}^{n+1} f_{K-1,i_{K-1}} (\lambda_{K-1,i_{K-1}|K} - \lambda_{K-1,i_{K-1}|K-1}) \\
    &= p^{\ast}(\bm{f}_{1:K-1}) \\
    &\ \ \ \ + (\bm{\lambda}_{1|K-1} - \bm{\lambda}_{1|K})'\bm{\mu}_{1} + \cdots +(\bm{\lambda}_{K-1|K-1} - \bm{\lambda}_{K-1|K})' \bm{\mu}_{K-1}
    + \bm{\lambda}_{K|K} ' (\bm{f}_{K} - \bm{\mu}_{K}) \\ 
    &\ \ \ \ + (\bm{\lambda}_{1|K} - \bm{\lambda}_{1|K-1})' \bm{f}_1 +\cdots +  (\bm{\lambda}_{K-1|K} - \bm{\lambda}_{K-1|K-1})' \bm{f}_{K-1}  \\
    &= p^{\ast}(\bm{f}_{1:K-1}) \\ 
    &\ \ \ \ - [ \ p + \bm{\lambda}_{1|K-1}'(\bm{f}_1 - \bm{\mu}_1) + \cdots + \bm{\lambda}_{K-1|K-1}'(\bm{f}_{K-1} - \bm{\mu}_{K-1}) \ ] \\
    &\ \ \ \ + [ \ p + \bm{\lambda}_{1|K}'(\bm{f}_1 - \bm{\mu}_1) + \cdots + \bm{\lambda}_{K-1|K}'(\bm{f}_{K-1} - \bm{\mu}_{K-1}) + \bm{\lambda}_{K|K}'(\bm{f}_{K} - \bm{\mu}_{K}) \ ] \\
    &= p^{\ast}(\bm{f}_{1:K-1}) -  p^{\ast}(\bm{f}_{1:K-1})  + p^{\ast}(\bm{f}_{1:K}) \\
    &=  p^{\ast}(\bm{f}_{1:K}). 
    \end{split}
\end{equation*}
This completes the proof by induction.

\subsection{Proof by using independent prior} \label{sec:indep}

The Consistency Condition requires {\it any} prior whose marginal means are $\bm{\mu}_{1:K}$ to satisfy the same equations. If we specifically consider the following independent prior, 
\begin{equation*}
    q(\bm{f}_{1:K}) \equiv \prod _{k=1}^K q(\bm{f}_k),
\end{equation*}
where $q(\bm{f}_k)$ has mean $\bm{\mu}_k$, then equation~(\ref{eq:linear_pool_multi}), or (\ref{eq:linear_pool_k}), must hold even for this prior. Our second proof is based on this fact, and uses the independent prior to derive the expression in (\ref{eq:discrete_bps}). 

Using the independence, $q( \bm{f}_K | \bm{f}_{1:K-1} ) = q(\bm{f}_K)$, and the Consistency Condition, we have 
\begin{equation*}
    p^{\ast}(\bm{f}_{1:K-1})  = \int p^{\ast}(\bm{f}_{1:K}) q(\bm{f}_K) \ d\bm{f}_K.
\end{equation*}
Relating this result to (\ref{eq:linear_pool_k}), we have 
\begin{equation*}
    \begin{split}
      p^{\ast}(\bm{f}_{1:K-1}) &= p + \bm{\lambda}_{1|K-1}' (\bm{f}_1 - \bm{\mu}_1) + \cdots + \bm{\lambda}_{K-1|K-1}' (\bm{f}_{K-1} - \bm{\mu}_{K-1}) \\
     &= p + \bm{\lambda}_{1|K}' (\bm{f}_1 - \bm{\mu}_1) + \cdots + \bm{\lambda}_{K-1|K}' (\bm{f}_{K-1} - \bm{\mu}_{K-1}),
    \end{split}
\end{equation*}
which shows $\bm{\lambda}_{k|K-1} = \bm{\lambda}_{k|K}$ for any $k \le K$, meaning that the coefficients of the $k$-th agent is invariant even if other agents are added to, or omitted from, the information set. We write $\bm{\lambda}_{k} \equiv \bm{\lambda}_{k|k'}$ for $k' \in \{ k,\dots, K \}$. Since equation~(\ref{eq:linear_pool_multi}), or (\ref{eq:linear_pool_k}), holds for any prior, coefficient $\bm{\lambda}_{k}$ does not depend on the number of agents even when we use priors other than $q(\bm{f}_{1:K})$. 

Now, let the DM's prior be an arbitrary $p(\bm{f}_{1:K})$ (but with mean $\bm{\mu}_{1:K}$). The result on coefficients greatly simplifies (\ref{eq:key}) as, 
\begin{equation*}
    \sum_{i_k = 1}^{n+1} \mathbb{P}(E|f_{1,i_1} = 1 , \cdots , f_{k,i_k} = 1)f_{k,i_k} = \mathbb{P}(E|f_{1,i_1} = 1 , \cdots , f_{k-1,i_{k-1}} = 1) + \bm{\lambda}_{k} ' (\bm{f}_{k} - \bm{\mu}_{k}).
\end{equation*}
Using this equation recursively, we obtain 
\begin{equation*}
\begin{split}
    & \sum_{i_1 = 1}^{n+1}\cdots \sum_{i_{K-1} = 1}^{n+1} \sum_{i_K = 1}^{n+1} \mathbb{P}\left( E \left| \  f_{1,i_1} = 1 , \cdots,f_{K,i_{K}} = 1 \right. \right) f_{1,i_1}\cdots f_{K,i_K}  \\
    &= \sum_{i_1 = 1}^{n+1}\cdots \sum_{i_{K-1}= 1}^{n+1}f_{1,i_1}\cdots f_{K-1,i_{K-1}}
    \sum_{i_K = 1}^{n+1}   \mathbb{P}\left( E \left| \  f_{1,i_1} = 1 , \cdots,f_{K,i_{K}} = 1 \right. \right) f_{K,i_K} \\
    &= \sum_{i_1 = 1}^{n+1}\cdots \sum_{i_{K-1}= 1}^{n+1} \mathbb{P}\left( E \left| \  f_{1,i_1} = 1 , \cdots,f_{K-1,i_{K-1}} = 1 \right. \right)  f_{1,i_1}\cdots f_{K-1,i_{K-1}} 
    + \bm{\lambda}_K' ( \bm{f}_K-\bm{\mu}_K ) \\
    &= \cdots \\
    &= \sum_{i_1 = 1}^{n+1} \mathbb{P}\left( E \left| \  f_{1,i_1} = 1 \right. \right)f_{1,i_1} + \bm{\lambda}_2' ( \bm{f}_2-\bm{\mu}_2 )  + \cdots + \bm{\lambda}_K' ( \bm{f}_K-\bm{\mu}_K )  \\
    &= p + \bm{\lambda}_1' ( \bm{f}_1-\bm{\mu}_1 ) + \bm{\lambda}_2' ( \bm{f}_2-\bm{\mu}_2 )  + \cdots + \bm{\lambda}_K' ( \bm{f}_K-\bm{\mu}_K ),
\end{split}
\end{equation*}
where the last line follows from the result for the single agent, or (\ref{eq:wc1}). The last expression is exactly $p^{\ast}(\bm{f}_{1:K})$ in (\ref{eq:linear_pool_k}). This completes the proof of (\ref{eq:discrete_bps}).

\subsection{Remarks}

\subsubsection*{On prior independence}

It is worth emphasizing that we do {\it not} assume the independence of agents in the DM's prior. Rather, the prior independence of agents {\it might} be true, but is not certain. The Consistency Condition must be satisfied in any case, regardless of the prior independence. This allows us to use the independent prior in the proof to compute the posterior probability, because it must be identical for any choice of priors. 

If the DM does assume the independence across agents--- if the DM excludes any other priors but decides to use $q(\bm{f}_{1:K})$ only--- then such an assumption must be considered explicitly when developing the predictive synthesis equation. In this case, the Consistency Condition must be modified, and the DM obtains a different form of updating rule \citep[Theorem~4.1]{genest1985modeling}.

\subsubsection*{Random number generation and independence}

To generate a random variable $y$ from the DM's posterior $p(y|\mathcal{H})$ in (\ref{eq:bps}), one should generate $x_k$ from each agent forecast distribution $h_k(x_k)$ independently, then generate $y$ from $\alpha (y|\bm{x})$. Similarly, to generate a binary variable whether event $E$ occurs or not from the DM's posterior $p^{\ast}(\bm{f}_{1:K})$ in (\ref{eq:discrete_bps}), one should generate index $i_k \in \{ 1,\dots ,n,n{+}1 \}$ from the discrete distribution with probability $\bm{f}_k$ for each $k$ independently, then generate a binary variable from the Bernoulli distribution with probability $\Pi _{(i_1,\dots, i_K)}$. 

The step of generating $x_k$ or $i_k$ independently is due to the product of agent forecast densities $h_k(x_k)$ in (\ref{eq:bps}), or the product of agent forecast probabilities $f_{k,i_k}$ in (\ref{eq:discrete_bps}). However, for the same reason stated in the previous remark, this does not mean that we assume the independence of agents. These products appear in the probability of event $E$ where the agent forecasts are conditioned. The (in)dependence of agents is stated in the DM's prior $p(\bm{f}_{1:K})$, but we avoided specifying this prior completely. To reiterate, we did specify its mean $\bm{\mu}_{1:K}$, not the (in)dependence.

\subsubsection*{Further extensions: multivariate forecasts by agents}

For simplicity, assume $K=1$ in this subsection. In proving (\ref{eq:bps}) by using (\ref{eq:wc1}), we assumed that the agent forecasts a single quantity $X$. The extension to the multivariate case, where $p$ quantities $\bm{X}=(X_1,\dots ,X_p)'$ are forecast by the agent, is straightforward. Consider the same equal-length grids in Section~\ref{sec:single_multi}, then define, 
\begin{equation*}
    \Delta _{i_1,\dots , i_p} \equiv  (q_{i_1-1}, q_{i_1}] \times \cdots \times (q_{i_p-1}, q_{i_p}] \subset \mathbb{R}^p,
\end{equation*}
where $i_1,\dots ,i_p \in \{ 1,\dots ,n\}$ and each interval has length $\epsilon _n>0$. Note that, for $i_1,\dots ,i_p \in \{ 2,\dots ,n{-}1\}$, the $\Delta_{i_1,\dots , i_p}$ is the $p$-dimensional hypercube with volume $\epsilon_n^p$. If we set $E_{i_1,\dots , i_p} = \{ \bm{X} \in \Delta_{i_1,\dots , i_p} \}$, then this problem becomes the synthesis of ``single agent, $np$ events.'' The synthesis function implied by (\ref{eq:wc1}) is,
\begin{equation*}
    \mathbb{P}( Y\le y | \mathcal{H}_n ) = \sum _{i_1=1}^n \cdots \sum _{i_p=1}^n \Pi _{i_1,\dots ,i_p}(y) H\!\left( \bm{X} \in \Delta_{i_1,\dots , i_p} \right),
\end{equation*}
where $H$ is the agent's distribution function of $\bm{X}$, and the coefficient is given by
\begin{equation*}
    \Pi _{i_1,\dots ,i_p}(y) \equiv \mathbb{P}( Y\le y | \bm{X} \in \Delta _{i_1,\dots ,i_p}). 
\end{equation*}
Viewing this expression as the Riemann sum, letting $n\to\infty$ and $\epsilon_n\to 0$ and computing the derivatives in $y$, we have
\begin{equation*}
    p( y | \mathcal{H} ) = \int _{\mathbb{R}^p} \alpha (y|\bm{x}) h( \bm{x} ) d\bm{x},
\end{equation*}
where $h$ is the density function of distribution $H$. The similar expression can be found for $K\ge 1$ in the same way. 

In the discussion above, we have implicitly assumed that the DM has a prior for $h$, or the agent's joint distribution of $\bm{X}$, and specifies its mean. If the agent does not report the joint density $h(\bm{x})$, but the the marginal densities, $h_1(x_1),\dots ,h(x_p)$, the synthesis formula becomes 
\begin{equation*}
    p( y | \mathcal{H} ) = \int _{\mathbb{R}^p} \alpha (y|\bm{x}) \prod _{j=1}^p h_j(x_j) d\bm{x}.
\end{equation*}
This synthesis formula also applies to the case where the agent reports joint density $h$, but the DM specifies the means of marginals $(h_1,\dots h_p)$, not the mean of joint $h$. Such a scenario has been considered in \cite{mcalinn2021mixed}.

\section*{Acknowledgments}
We thank Mike West for his comments and suggestions. 
The second author's research was partly supported by JSPS KAKENHI Grant Number 22K13374, Japan Society for the Promotion of Science. This research was conducted while the first author was a graduate student at the Graduate School of Economics, the University of Tokyo. Any opinions expressed in this article are those of the authors.

\bibliographystyle{apalike}
\bibliography{bayes}

\end{document}